# Optimal SMF packing in photonic lanterns: comparing theoretical topology to practical packing arrangements

JOHN J. DAVENPORT,[1] MOMEN DIAB,[1] KALAGA MADHAV,[1] MARTIN M. ROTH[1]

[1]*Leibniz-Institut für Astrophysik Potsdam (AIP), An der Sternwarte 16, D-14482 Potsdam, Germany*

**Abstract:** Photonic lanterns rely on a close packed arrangement of single mode fibers, which are tapered and fused into one multi-mode core. Topologically optimal circle packing arrangements have been well studied. Using this, we fabricate PLs with 19 and 37 SMFs showing tightly packed, ordered arrangements with packing densities of 95 % and 99 % of theoretically achievable values, with mean adjacent core separations of 1.03 and 1.08 fiber diameters, respectively. We demonstrate that topological circle packing data is a good predictor for optimal PL parameters.

## 1. Introduction

Photonic lanterns (PLs) are fiber optic or photonic devices, for transitioning between a multi mode fiber (MMF) and a number of single mode fibers (SMFs), a multi-core fiber (MCFs) or a series of waveguides [1, 2, 3]. They have a wide range of applications in astrophotonics such as mode scrambling of optical signals [4, 5, 6], reformatting light into a pseudo-slit [7, 8, 9], and for introducing complex fiber Bragg gratings [10, 11, 12] in fiber-fed spectrographs for suppressing sky OH-emission in ground-based astronomy, and space-division multiplexing (SDM) [13, 14, 15] in telecommunications.

A common form of PL consists of a bundle of SMFs packed inside a low refractive index capillary [1, 3, 16]. Heat and axial tension are then applied to taper down one end of the arrangement, where the claddings of the individual SMFs fuse together to form the core of the MMF, while the capillary becomes the new cladding of the MMF. Low refractive index capillaries such as fluorine doped silica are used to contain light within the new core [1].

Fontaine et. al. [17] carried out simulations based on modal analysis of PLs to investigate the geometric requirements of SMF arrangements in PLs when transitioning between SMF and MMF regions. It was found that efficient coupling is strongly dependent on arrangement of cores, with best results achieved when supermodes supported within the SMFs close to the transition region approximate to the modes supported in the MMF region. In practice, this generally requires ordered, controllable arrangements of SMFs with minimal gaps between cores.

However, the SMF taper method described above does not allow for manual control of SMF arrangement. Several authors have found that ordered arrangements can only be obtained by careful matching of the capillary inner diameter (OD) with number and diameters of the SMFs, mostly using seven or fewer SMFs [3, 18, 19] . The requirement for efficient packing of SMFs within the capillary can then force them into an ordered arrangement. Noordegraaf et al. [20] demonstrated an impressive PL with 61 SMFs.

Other methods include tapering of an MCF inside a solid capillary [4, 5], direct etching into a glass substrate using ultrafast laser inscription (ULI) [21], and hole collapse of a photonic crystal fiber (PCF) [1]. This paper focuses primarily on the first method, for which the core layout cannot be freely controlled during manufacturing. For example it is common for the cores of MCFs to be arranged in a hexagonal pattern during production of the preform. However, cores or waveguides in the other methods mentioned above could be intentionally placed in patterns described here if efficient packing within a circular arrangement was desired.

In the mathematical field of topology, the question of how congruent circles pack within a larger circle has been well studied. Graham [22] proved the optimal packing arrangements for up to 7 circles. Pirl et al. [23] expended this list up to 19 circles, with the case of 19 proved later by Fodor [24], and arrangements up to 25 were found by Reis et al. [25]. Graham et al. [26] used two packing algorithms to find optimal arrangements up to 65, selecting the best (densest) packed result in each case. However, Graham et. al. report that the two algorithms found the same best packing result in the majority of cases.

In this paper we make use of the best packing arrangements found by Graham et. al. to predict and control packing of SMFs within a capillary. We fabricated PLs with different SMF numbers and capillary ODs. By cleaving the PL at the transition regions, the resulting SMF arrangements were analyzed. The fabricated PLs showed arrangements that were well matched to the theoretical optimal arrangements, high packing densities, and low core separation distances. This work may additionally be useful for other devices where close packing of optical fibers is desirable, such as imaging fiber arrays and wavefront sensors [27, 28].

## 2. Theory

In topology, the question of the most efficient packing of circles in a larger circle can be expressed in a number of ways. We ask the following question: given $n$ congruent (equal size and identical shape) nonoverlapping circles, which arrangement allows them to be placed entirely inside a larger circle with the minimum radius? The corresponding arrangement is called an optimal packing [26]. This is a 2D simplification of a 3D SMFs inside a glass capillary, where the congruent circles are the individual SMFs, and inner diameter (ID) of the capillary is the enclosing circle.

The ratio between the diameters of the inner circles and outer circle is given by:

$$R = \frac{d_1}{d_2}, \qquad (1$$

where $d_1$ is the diameter of the outer circle (ID of the capillary), and $d_2$ is the diameter of the inner circles (SMFs). The packing density can then be found as the sum of the areas of the inner circles, over the area of the outer circle:

$$D_n = \frac{n\,\pi\left(\frac{d_2}{2}\right)^2}{\pi\left(\frac{d_1}{2}\right)^2},$$

which can be simplified to:

$$D_n = n\left(\frac{d_2}{d_1}\right)^2, \qquad (2$$

Graham et al. [26] found packing arrangements for circles up to $n = 65$, using two types of packing methods. One was ordinary non-linear optimization algorithms with an approximate cost function, the other simulated the movement of idealized billiard balls inside a circular container. The full list of arrangements can be found in [26], and several arrangements are presented here in Fig. 1. These specific arrangements represent a nonexhaustive selection of stable and unique arrangements with high packing density relative to neighboring numbers, which were deemed desirable criteria when producing PLs.

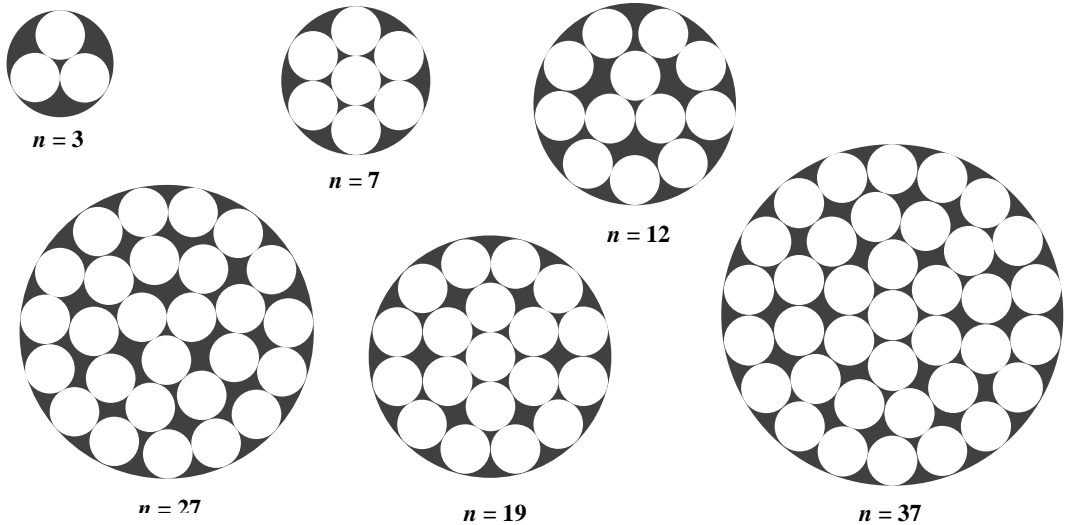

Fig. 1. Optimal packing arrangements for congruent circle numbers: 3, 7, 12, 19, 27, 37. [26].

Table 1 shows the diameter of the outer circle, relative to the inner circles, calculated from [26] and Eq. (2). Data is given in this form for convenience in selecting or tapering a capillary to fit a particular size and number of SMFs. Arrangements were then ranked as their recommended suitability for use as a PL, as very good, good, okay and bad. These recommendations are made considering high density relative to neighbors, the presence of regular concentric rings, rings containing odd numbers, and stability and uniqueness of arrangement. The reasons for each of these considerations are given in the following sections. However, it is important to note that these recommendations remain subjective judgements. Fig. 2 is a graph of packing densities, calculated from data in Table 1. In general, higher density relative to neighbors is indicative of an efficient, well ordered and tight arrangement of SMFs, but there are a number of exceptions as shown in Table 1. It is noticeable that density peaks can be seen at the ring hexagonal numbers of 7, 19, 37 and 61. A density peak can also be seen at $n$=55, which happens to show an unusually dense packing.

**Table 1. Optimal outer circle diameters (capillary IDs), relative to inner circle diameter (AU), along with comments and recommended rankings of arrangements for use in PL fabrication.**

| $n$ | $d_1$ (AU) | comment | Rank | $n$ | $d_1$ (AU) | comment | Rank |
|---|---|---|---|---|---|---|---|
| 2 | 2.000 | | Good | 34 | 6.610 | 4 rings, unstable, irregular | Okay |
| 3 | 2.154 | 1 ring, high density | Very good | 35 | 6.697 | 4 rings, unstable, irregular | Okay |
| 4 | 2.414 | 1 ring, high density | Good | 36 | 6.746 | 4 rings, unstable, irregular | Okay |
| 5 | 2.701 | 1 ring | Good | 37 | 6.758 | 4 rings, high density | Very good |
| 6 | 3.000 | Unstable, 2 arrangements | Bad | 38 | 6.961 | 4 rings, unstable, irregular | Okay |
| 7 | 3.000 | 2 rings, high density | Good | 39 | 7.057 | 4 rings, unstable, irregular | Okay |
| 8 | 3.304 | 2 rings, unstable | Okay | 40 | 7.123 | 4 rings, all odd, unstable | Good |
| 9 | 3.613 | 2 rings, unstable | Okay | 41 | 7.260 | Unstable, irregular | Bad |
| 10 | 3.813 | 2 rings, irregular | Okay | 42 | 7.346 | Unstable, irregular | Bad |
| 11 | 3.923 | 2 rings, 2 arrangements, irregular | Bad | 43 | 7.419 | Unstable, irregular | Bad |
| | | | | 44 | 7.498 | Unstable, irregular | Bad |
| 12 | 4.029 | 2 rings | Good | 45 | 7.576 | 4 rings, unstable, irregular | Bad |
| 13 | 4.236 | 2 rings, irregular | Okay | 46 | 7.650 | 4 rings, unstable, irregular | Okay |
| 14 | 4.328 | 2 rings, irregular | Okay | 47 | 7.724 | 4 rings, unstable, irregular | Okay |
| 15 | 4.521 | 2 rings | Good | 48 | 7.791 | 4 rings, unstable | Good |
| 16 | 4.615 | 2 rings | Good | 49 | 7.886 | Unstable, irregular | Bad |
| 17 | 4.792 | 3 rings, irregular | Okay | 50 | 7.947 | Unstable, irregular | Bad |
| 18 | 4.863 | Unstable, 10 arrangements | Bad | 51 | 8.027 | 4 rings, unstable | Okay |
| 19 | 4.863 | 3 rings, high density | Very good | 52 | 8.084 | 4 rings, unstable | Okay |
| 20 | 5.122 | 3 rings, unstable, irregular | Okay | 53 | 8.179 | 5 rings, unstable, irregular | Bad |
| 21 | 5.252 | 3 rings, unstable | Okay | 54 | 8.203 | 5 rings, unstable, irregular | Okay |
| 22 | 5.439 | Unstable, irregular | Bad | 55 | 8.211 | 5 rings, unstable | Good |
| 23 | 5.545 | 3 rings, irregular | Okay | 56 | 8.384 | Unstable, irregular | Bad |
| 24 | 5.651 | 3 rings, unstable, irregular | Bad | 57 | 8.447 | Unstable, irregular | Bad |
| 25 | 5.752 | 3 rings, unstable | Okay | 58 | 8.524 | 5 rings, unstable, irregular | Okay |
| 26 | 5.828 | Unstable, irregular | Bad | 59 | 8.592 | 5 rings, unstable, irregular | Okay |
| 27 | 5.906 | 3 rings | Very good | 60 | 8.646 | 5 rings, unstable, irregular | Okay |
| 28 | 6.014 | 3 rings, unstable, irregular | Okay | 61 | 8.661 | 5 rings, 3 arrangements, high density | Good |
| 29 | 6.138 | Unstable, irregular | Bad | 62 | 8.829 | 5 rings, unstable, irregular | Okay |
| 30 | 6.197 | 3 rings | Good | 63 | 8.892 | 5 rings, unstable, irregular | Okay |
| 31 | 6.291 | 3 rings, gaps around edge | Okay | 64 | 8.961 | Unstable, irregular | Bad |
| 32 | 6.429 | 3 rings, unstable | Okay | 65 | 9.017 | 5 rings, unstable, irregular | Okay |
| 33 | 6.486 | 3 rings | Good | | | | |

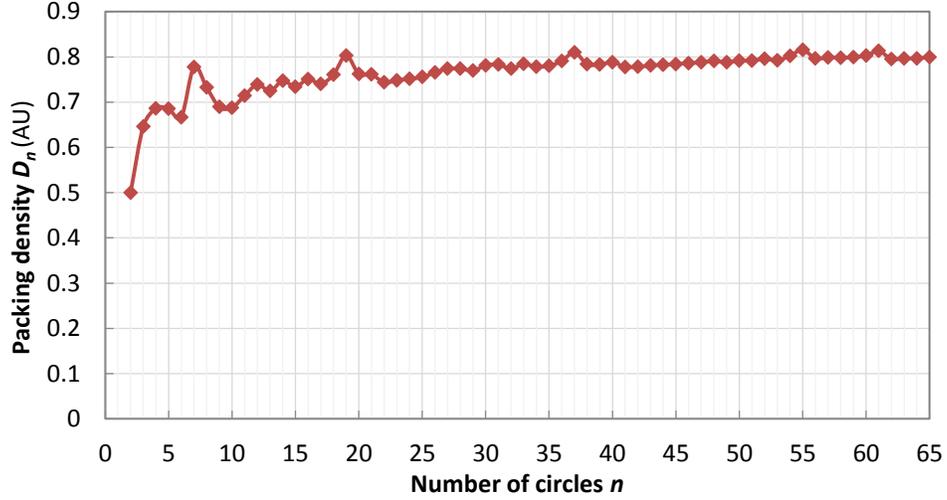
Fig. 2. Graph of the packing densities of optimal arrangements, $D_n$.

### 2.1 Photonic lantern mode coupling

Optimal coupling efficiency in a PL can only be achieved when the number of modes supported by the output region is equal to or greater than the number of modes supported by the input region [1]. The approximate number of unpolarized modes supported within an MMF is given by:

$$N_{mm} \approx \left(\frac{V_{mm}}{2}\right)^2 = \left(\frac{\pi d_{mm} NA_{mm}}{2\lambda}\right)^2, \qquad (3)$$

where $V_{mm}$ is the V-parameter for the fiber [1], $d_{mm}$ is the diameter of the fiber core, $NA_{mm}$ is the numerical aperture and $\lambda$ is the wavelength. Assuming all the cores in the SM region support a single, unpolarized mode, the number of supermodes supported by the SMF region, $N_{sm}$, is equal to the number of SMFs. PLs can be used in either MMF to SMF or SMF to MMF directions.

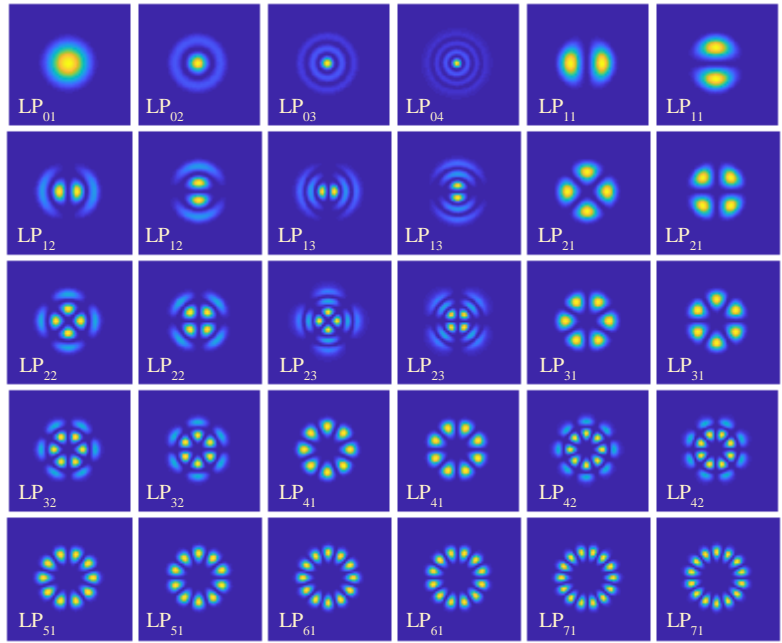
Fig. 3. Modal geometries of a 30 mode MMF.

A number of studies have theoretically demonstrated that best coupling can be found between the two regions when the SMF arrangement approximates or samples the geometry of MMF modes [17, 29]. In practice the exact coupling between regions is very complicated and will be influenced by perturbations and imperfections in the fibers, intensity and phase of each mode etc., and the modal geometries are unique to each PL [2], but a number of similar patterns can be observed. Fig. 3 shows simulated modal geometries of an MMF, featuring 30 modes. Modes of this geometry are described as $LP_{lm}$, where *l* is the maximum azimuthal index and *m* is the radial node number [2].

Leon-Saval et. al. [2] says that the best coupling can be achieved when there are m concentric rings, and each ring contains 2*p*+1 SMFs, where *p* is the largest present *l* for each *m*. An odd number of SMFs in each ring is desirable to differentiate between the cos(*l*φ) and sin(*l*φ) azimuthal modal dependencies, but this is not always possible. In the example shown in Fig. 3, m varies from 1 to 3 with the exception of the final mode $LP_{04}$, and 2*p*+1 values from 1 to 15. The arrangement for 27 SMFs has 3 concentric rings with 15 SMFs in the outer ring, and would work well for an SMF to MMF coupling. The arrangement for 37 SMFs has 4 rings with 18 SMFs in the outer ring, suitable for MMF to SMF coupling. The arrangement for 30 SMFs also has 3 rings with 16 SMFs in the outer ring and matches the number of modes in the MMF, and is a good compromise between the two directions.

*2.2 Alternative and Unstable Arrangements*

Not all of the circle arrangements found in [26] can be used in practice for fabricating PLs. Fig. 4 shows some examples of arrangements not suitable for use in PLs. Circles in alternate packing solutions are analyzed by Lubachevsky and Graham [30]. For *n*=6 two types of arrangements are possible with the same ratio. Fig.4: 6a is stable with all 6 circles arranged around the rim, but is not suitable for PLs due to the large gap at the center. Fig. 4: 6b has one circle in the center and five arranged around the rim, where the individual circles along the circumference are in contact with only the center circle. For PLs, this arrangement cannot be packed more densely and is unstable. The gaps between the five outer circles may not hold equally during the drawing and tapering process, and it is likely that some of the fibers may fall to one side leaving a large air gap on the other. Some arrangements, such as n=40 and n=48, have a very small degree of instability which is unlikely to cause a problem in drawing a PL. For *n*=11, two arrangements are possible, Fig. 4: 11a and Fig. 4: 11b. Both are stable, but unusable due to unwanted gaps, and offers no control over which arrangement is formed during fabrication.

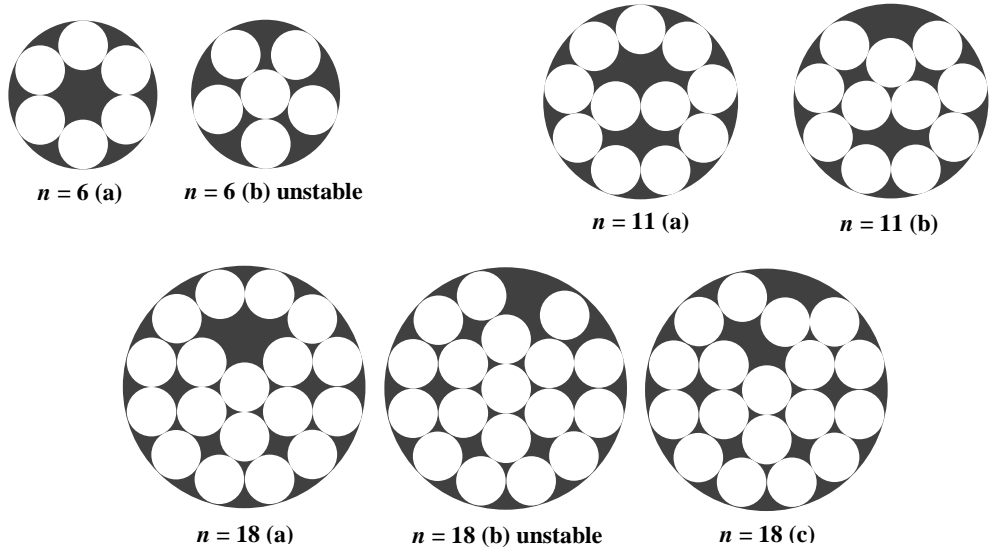

**n = 6 (a)**   **n = 6 (b) unstable**   **n = 11 (a)**   **n = 11 (b)**

**n = 18 (a)**   **n = 18 (b) unstable**   **n = 18 (c)**

Fig. 4. Unusable examples of packing arrangements for PLs for *n*= 6, 11 and 18.

For *n*=18 circles, ten different, rotationally unique arrangements were shown in [26], a selection of which are shown in Fig. 4. *n*=18 circles has the same *R* as *n*=19, and several of the arrangements are the same as for *n*=19 circles with one missing circle. All but one of these arrangements are stable (Fig. 4: 18b is unstable), but as with *n*=11, there is no control over which arrangement is formed. *n*=61, not shown in Fig. 4, offers three different arrangements, where all three are stable and very similar, essentially just rotations of the outer rings. When designing PLs, it is important to be aware that while most numbers have only a single arrangement, those listed here have more than one and are sometimes unstable.

## 3. Method

The PL was produced as described in Birks et. al. [1]. A bundle of SMFs were packed inside a low refractive index capillary which was then heated and tapered to produce a single MMF. The tapering was performed using a Vytran GPX-3000 glass processing system. The SMFs used were F-sm1500-6.4/80 and F-sm1250-9/80 from Newport Corporation, with pure silica cladding of 80 μm diameter. Taper ratio is important for PLs coupling light from SMFs to MMF: high taper ratio can result in light leaking out of the transition region into the surrounding capillary to become cladding modes [1]. Adiabatic taper length (L) scales approximately with $L \propto n^2$ [19]. As the core of the MMF was formed from the claddings of the SMFs, fluorine doped capillaries were used to provide a lower refractive index cladding around this core. Fluorine doped capillaries were manufactured and provided by the Centre for Photonics and Photonic Materials, University of Bath.

In order to pack SMFs tightly within a capillary, the ID of each low index capillary has to match the minimum outer circle diameter described in the theory section. Capillaries with larger IDs were therefore selected, which were tapered down to the desired ID prior to packing with SMFs. This had the added advantage that the down-taper of the capillary formed a funnel shape that helped collect the SMFs and force them into an ordered arrangement. The final ID of the tapered capillary was larger than the theoretical ID by few μm. This is to ensure that the SMFs do not to jam during the packing phase. Fig. 5 shows a schematic of the tapered capillary with SMFs. Capillaries were then tapered and drawn down, to form PLs.

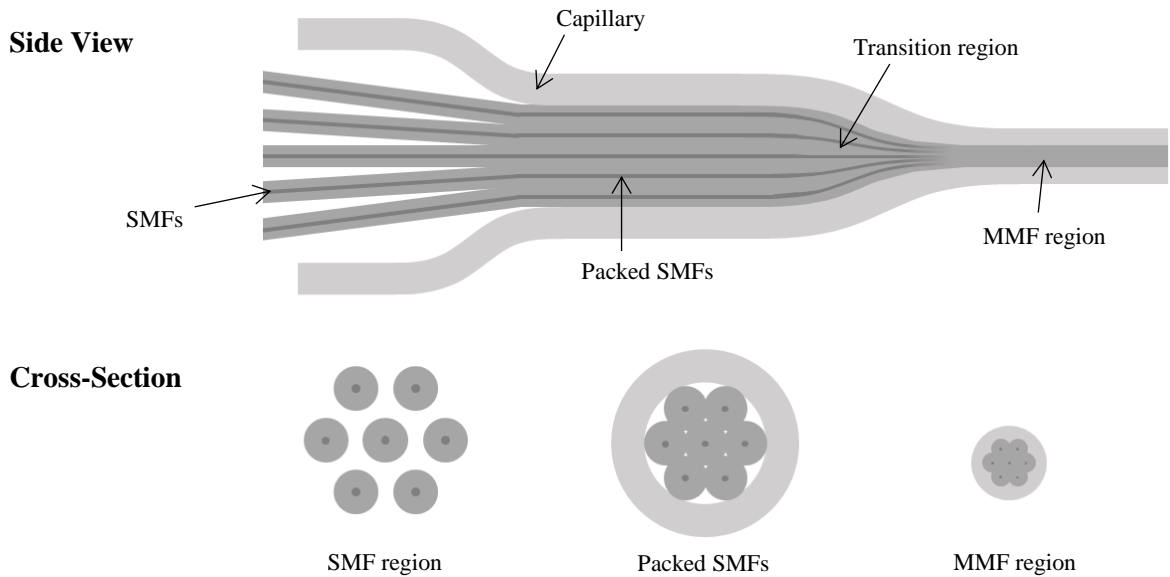

Fig. 5. Schematic showing the tapering process of an example PL for *n*=7

## 4. Results

The arrangements of SMFs in the PLs were compared to the optimal circle packing arrangements by cleaving the fiber bundle at the taper section after drawing, and imaging the cross-section with a microscope. Analysis of the taper section is particularly important for finding the positions of the SMF cores, as by design these cores become too small closer to the MMF section to efficiently couple light after tapering. Fig. 6 and 7 show the comparison of theoretical packing to the fabricated PLs at the taper sections for *n*=19 and 37 SMFs, respectively.

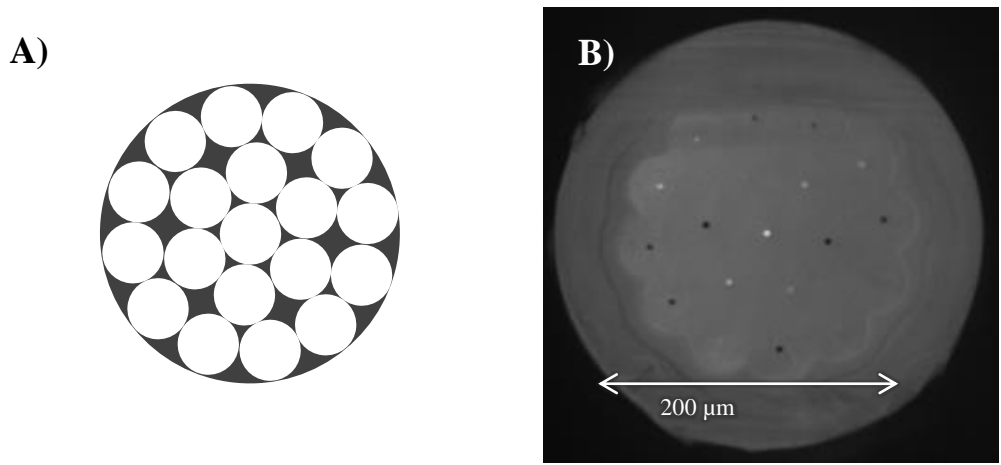

Fig. 6. A) Optimal theoretical packing arrangement for *n*=19 congruent circles. B) Micrograph of a transition region cleave of an *n*=19 SMF PL.

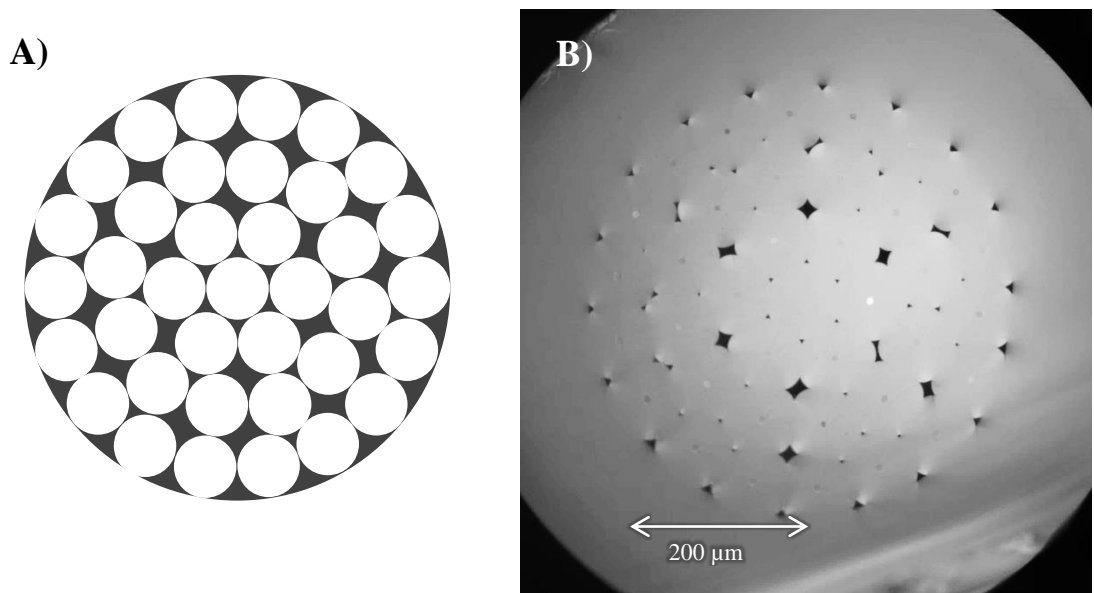

Fig. 7. A) Optimal theoretical packing arrangement for *n*=37 congruent circles. B) Micrograph of a transition region cleave of a *n*=37 SMF PLs.

In order to quantify the similarity between fabricated and theoretical packing, the relative positions of SMF cores were measured on the transition region micrograph images shown in Fig. 6 and 7. Circle center positions in theoretical packing were then fitted to the fabricated SMF core positions using a least-squares fit considering circle size (and therefore center point separation), x-y position and rotation. The results are shown in Fig. 8 and 9.

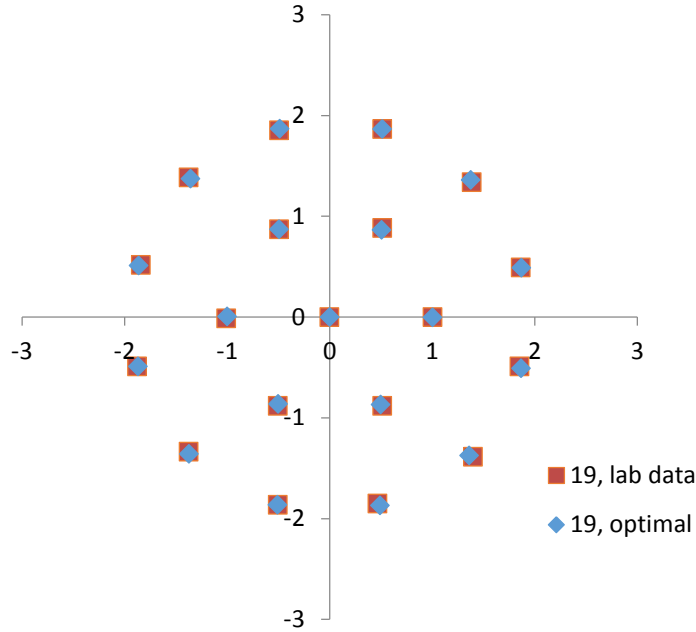

Fig. 9. Positions of SMF cores for $n$=19 SMF PL, compared to circle center positions of the theoretical optimal arrangement. Red squares represent laboratory data and blue rhombuses represent the optimal arrangement.

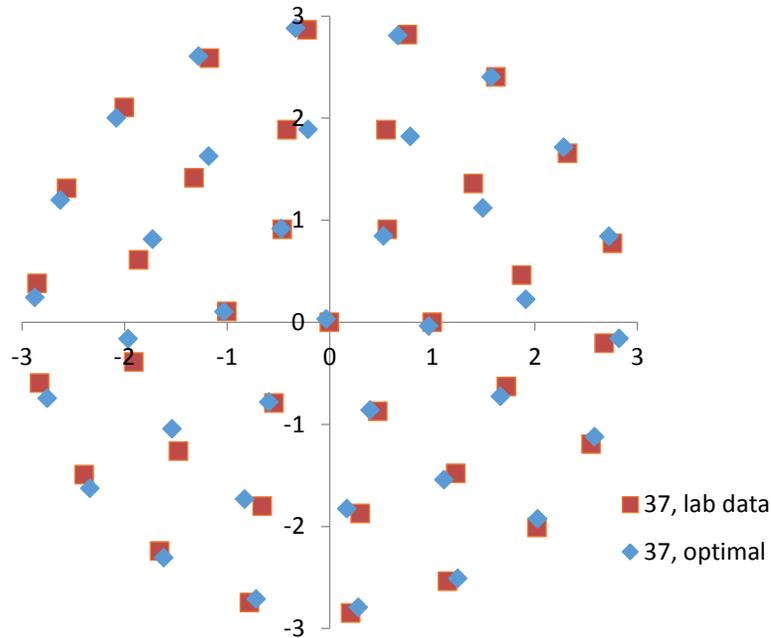

Fig. 10. Positions of SMF cores for $n$=37 SMF PL, compared to circle center positions of the theoretical optimal arrangement. Red squares represent laboratory data and blue rhombuses represent the optimal arrangement.

The $n$=19 SMF arrangement is very close to theoretical arrangement, with a mean average difference between observed and theoretical positions $\Delta P$=1.68 % of circle size and a standard deviation $\sigma$=1.04 %. The $n$=37 SMF arrangement shows the same basic form as the theoretical arrangement, as a series of concentric rings, but the outer rings show some relative rotation. The outermost and second outer rings are rotated by -1.87° and 6.01° relative to the theoretical arrangement, leading to an overall mean $\Delta P$=12.7 % and $\sigma$=7.10 %. Despite this both outer rings remain well ordered. When the rotation of the theoretical arrangement was considered for each ring in isolation, values for the outermost and second outer rings decreased to mean $\Delta P$=6.32 % and $\sigma$=2.88 %, and mean $\Delta P$=4.97 % and $\sigma$=2.98 %, respectively.

It is thought that it might be possible to remove this rotation with a tighter fit between the SMFs and the capillary, at the cost of making the insertion more difficult. This has a small effect on the packing density. Leon-Saval et. al. [2] describes mode coupling depending on number of rings and on SMFs within rings but does not reference a significance to orientation. However removing the unpredictability of this rotation may be advantageous.

Next, the mean distance between adjacent cores was found using the same transition region data. Cores in both PLs had between 3 and 6 adjacent cores, with mean separations of 1.03 and 1.08 fiber diameters for the $n=19$ and $n=37$ PLs, respectively. Packing densities were calculated from the pre-taper capillary and core diameters, giving 0.763 and 0.800, or 95% and 99% of topological densities given in Table 1. A summary of the analyzed details of the three PLs can be found in Table 2.

Table 2. Summary of details of the PLs drawn in this paper. Values are given relative to circle or SMF diameters unless stated otherwise.

| SMF number $n$ | Optimal Ø | Capillary ID | Density | Mean Core Separation | Mean Δ$P$ |
|---|---|---|---|---|---|
| 19 | 4.8637 | 4.99 | 0.763 | 1.03 | 1.68% |
| 37 | 6.7587 | 6.80 | 0.800 | 1.08 | 12.7% |
| | | | | | |
| | | | | $n=37$ SMFs, outermost ring only, +1.87° | 6.32% |
| | | | | $n=37$ SMFs, second outer ring only, -6.01° | 4.97% |

## 5. Conclusion

PLs produced by the SMF taper method require a close arrangement of SMFs to be packed inside a glass capillary before tapering, but with current methods this cannot be achieved manually. It is possible that automated precision fiber placement combined with carefully positioned coreless spacer fibers may make this possible in the future, but such methods have not yet been developed. The optimal arrangements of congruent circles inside a larger circle have been well studied in topology, and are analogous to a cross-section of SMFs within a capillary.

In this study two PLs were fabricated by packing SMFs inside low index capillaries. The capillaries were pre-tapered to give internal diameters close to optimal packing diameters given by topological studies, with between 3 and 4 µm added to prevent jamming. This was then heated and drawn under tension to fuse the SMFs into a single MMF core. By cleaving PLs in the transition region, the arrangement of SMF could be seen and analyzed.

Results showed that ordered arrangements of SMFs were achieved that were close to optimal arrangements. $n=19$ SMFs was the most similar, with mean deviation from optimal positions of 1.68 % of SMF cladding diameter. The $n=37$ SMF PL took an arrangement of concentric rings similar to the theoretical arrangement, although the outermost and the second outer rings were rotated by -1.87° and 6.01° respectively. It is thought that this rotation occurred due to the extra diameter added to the capillary diameter to prevent SMFs jamming. However, packing densities of the $n=19$ and $n=37$ fiber PLs were found to be 95 % and 99 % of theoretical values respectively, with the mean separation between adjacent cores of 1.03 and 1.08 fiber diameters.

Until this point, only a small range of $n$ values have been tested, and finding the appropriate diameter ratio can require time consuming trial and error. Here we demonstrate that well studied topological circle packing arrangements are a good predictor for fiber arrangements within a capillary and using these ratios is an effective way of achieving high packing densities. The relationship between topological packing theory and fabrication of PLs will allow efficient fabrication of PLs with large $n$. Values of $n$ up to 65 were considered and it is thought larger numbers may also be possible. The influence of diameter and core concentricity tolerance of each SMF on the packing efficiency and performance of PL will be studied in future. These have adverse effects on fiber fed astrophotonic instruments in ground-based astronomy.

**Acknowledgements**

This work was supported by BMBF grant 03Z22AN11 "Zentren für Innovationskompetenz: innoFSPEC", and by the Deutsche Forschungsgemeinschaft (DFG) through project 326946494, 'Novel Astronomical Instrumentation through photonic Reformatting'.